\documentclass[5p,twocolumn,times,number]{elsarticle}

\usepackage{graphicx}
\usepackage{amsmath}   

\begin{document}

\begin{frontmatter}

\title{Charge Transfer Properties Through Graphene for Applications in Gaseous Detectors}

\author[add1]{S.~Franchino}
\author[add1]{D.~Gonzalez-Diaz}
\author[add2]{R.~Hall-Wilton}
\author[add3]{R.~B.~Jackman}
\author[add1]{H.~Muller}
\author[add3]{T.~T.~Nguyen}
\author[add1]{R.~de~Oliveira}
\author[add1]{E.~Oliveri}
\author[add1,add2]{D.~Pfeiffer}
\author[add1,add2]{F.~Resnati\corref{cor}}
\ead{filippo.resnati@cern.ch}
\author[add1]{L.~Ropelewski}
\author[add3]{J.~Smith}
\author[add1]{M.~van~Stenis}
\author[add4]{C.~Streli}
\author[add1,add4]{P.~Thuiner}
\author[add1,add5]{R.~Veenhof}

\cortext[cor]{Corresponding author}

\address[add1]{CERN, European Organization for Nuclear Research, CH-1211 Geneva 23, Switzerland}
\address[add2]{ESS, European Spallation Source, Tunav\"agen 24, 223 63 Lund, Sweden}
\address[add3]{London Centre for Nanotechnology and the Department of Electronic and Electrical Engineering, University College London, 17-19 Gordon Street, WC1H 0AH, United Kingdom}
\address[add4]{Atominstitut, Technische Universit\"at Wien, Stadionallee 2, 1020 Vienna, Austria}
\address[add5]{Department of Physics, Uluda\u{g} University, 16059 Bursa, Turkey}

\begin{abstract}
Graphene is a single layer of carbon atoms arranged in a honeycomb lattice with remarkable mechanical and electrical properties.
Regarded as the thinnest and narrowest conductive mesh, it has drastically different transmission behaviours when bombarded with electrons and ions in vacuum.
This property, if confirmed in gas, may be a definitive solution for the ion back-flow problem in gaseous detectors.
In order to ascertain this aspect, graphene layers of dimensions of about $2\times2$~cm$^2$, grown on a copper substrate, are transferred onto a flat metal surface with holes, so that the graphene layer is freely suspended.
The graphene and the support are installed into a gaseous detector equipped with a triple Gaseous Electron Multiplier (GEM), and the transparency properties to electrons and ions are studied in gas as a function of the electric fields.
The techniques to produce the graphene samples are described, and we report on preliminary tests of graphene-coated GEMs.
\end{abstract}


\end{frontmatter}

\section{Introduction}
%Graphene is a semimetal~\cite{Neto:2009}, which conserves~\cite{Zhou:2013} its conducting properties when large areas are supported on a metal.
%This allows the non-intrusive integration of the material on gold or copper, which constitutes existing detector parts.
Graphene is the thinnest material to date, consisting of a single layer of carbon atoms arranged in aromatic rings forming a honeycomb-like structure.
Its extraordinary mechanical properties allow for a stable integration of freely suspended graphene over holes of tens of micrometers in diameter.
%This very high mechanical resistance allows also for impermeability to ions and atoms, while allowing transparency to electrons over certain ranges of energy.
The basic hexagonal ring has a bond length of 0.142~nm and an inner radius of 0.246~nm.
The $\pi$~bonds orthogonal to the lattice can be seen as a de-localised cloud of electrons which overlaps the hole in the hexagon.
This reduces the opening pore, yielding an effective diameter of 0.064~nm~\cite{Berry:2013}, much smaller than the van der Waals radius of most atoms.
%The $\pi$~bonds therefore should allow impermeability
Graphene should then be impermeable to atoms, molecules and ions~\cite{Schedin:2007} when they do not have enough energy to go through the electron cloud.
Experimentally, suspended graphene has been measured to withstand an irradiation dose up to approximately $10^{16}$~ions/cm$^2$ at tens of~keV energies~\cite{Morin:2012}.
Other experiments have shown that graphene is completely impermeable to helium atoms up to 6~atm~\cite{Bunch:2008}.
On the other hand, graphene has been shown to exhibit high transparency to electrons with energies ranging from tens of~keV up to 300~keV~\cite{Meyer:2007}.
Transmission of electrons with energies lower than 10~eV energy is strongly hinted by reflectivity measurements performed in~\cite{Hibino:2008}, and attributed to quantum tunneling.
This asymmetry in electron and ion transport makes the graphene interesting for applications in gaseous detectors.
When used to separate the ionisation/conversion from the amplification region, the graphene would ideally eliminate the ion back-flow, hence overcoming the classical limit of gaseous detectors operating at high particle fluxes~\cite{Mathieson:1986}.

Compared to vacuum, charge transfer properties through graphene in gas are not well known.
The first studies~\cite{Thuiner:2015} showed that defects in the graphene layer play an important role in the electron and ion transmission measurements.
Nevertheless, an almost defect-free triple layer graphene can be successfully transferred onto metal meshes.
This three-atom-thick structure is able to block the drifting ions down to the sensitivity of the measurements, letting a small fraction of the drifting electrons pass through.
In the following sections we present the status and the improvements obtained in the production of the samples, and the first results from the operation of a GEM~\cite{Sauli:1997} coated with graphene.

\section{Production of the samples}
We exploit two procedures to produce suspended graphene over several tens of micrometers.
The first one involves the production of graphene onto a substrate followed by the transport of the graphene onto the final substrate.
The second one consists in directly producing the holes with photolithographic techniques into the substrate where the graphene is grown onto~\cite{O'Hern:2011}.
%Graphene is prepared in mono-layer, dual-layer, and tri-layer forms.
%The three layers allow the preparation of defect-free layers on copper meshes.

Concerning the first technique, the graphene is grown as a monolayer on polycrystalline copper by chemical vapour deposition (CVD) at 1000$^\circ$~C with CH$_4$ as a carbon precursor, and H$_2$ and Ar mixtures.
%The graphene grown on the back of the copper is then removed by immersion in nitric acid.
A layer of 300~nm of PMMA is spin coated onto the surface of graphene.
The copper support is removed using an aqueous solution of Fe(NO$_3$)$_3$.
The graphene attached to the remaining layer of PMMA is transferred onto the final substrate.
The PMMA is removed with acetone using a critical point dryer to avoid damages of the free-standing layer due to the surface tension of the solution.
Multi-layer graphene is made using several monolayers transferred one on top of each other and layered onto the final substrate.
The quality of the graphene is evaluated with Raman spectroscopy, and the coverage of the graphene is assessed using scanning electron microscopy (SEM).
The coverage of single layer is found to be 90$\pm$5\% and of triple layer is 99$\pm$0.5\%~\cite{Thuiner:2015}.
The advantage of this technique is that the graphene can be transferred onto any metal substrate, for instance one of the faces of a GEM.

Nevertheless, during the transfer and manipulation the graphene can be damaged.
For this reason we investigated a second approach that does not require the transfer at all.
Similarly to the previous technique, the basic material is CVD graphene grown onto a copper substrate.
The graphene is protected firstly with PMMA, and then with liquid photoresist.
Photosensitive photoresist applied on the face without graphene is selectively removed following the UV exposure, development and chemical etching procedures.
This process leaves unprotected copper discs of 50~$\mu$m in diameter at a distance of 140~$\mu$m.
The copper is chemically removed in these regions down to the graphene.
The final steps consist in the removal of the liquid photoresist with ethyl alcohol and the removal of the PMMA with acetone in the critical point dryer.
The result of the entire procedure is a copper grid, consisting of holes tapped with the graphene layer.
In spite of the fact that this technique is very new and not mature, encouraging preliminary results showed that what is left on the holes is pure single layer graphene.
The coverage is still not as good as the one achieved with the more established method of transferring the graphene.
The GEM sample used in the tests described next was obtained with the first method.

\section{Experimental apparatus}
\begin{figure}
\centering
\includegraphics[width=1.0\linewidth]{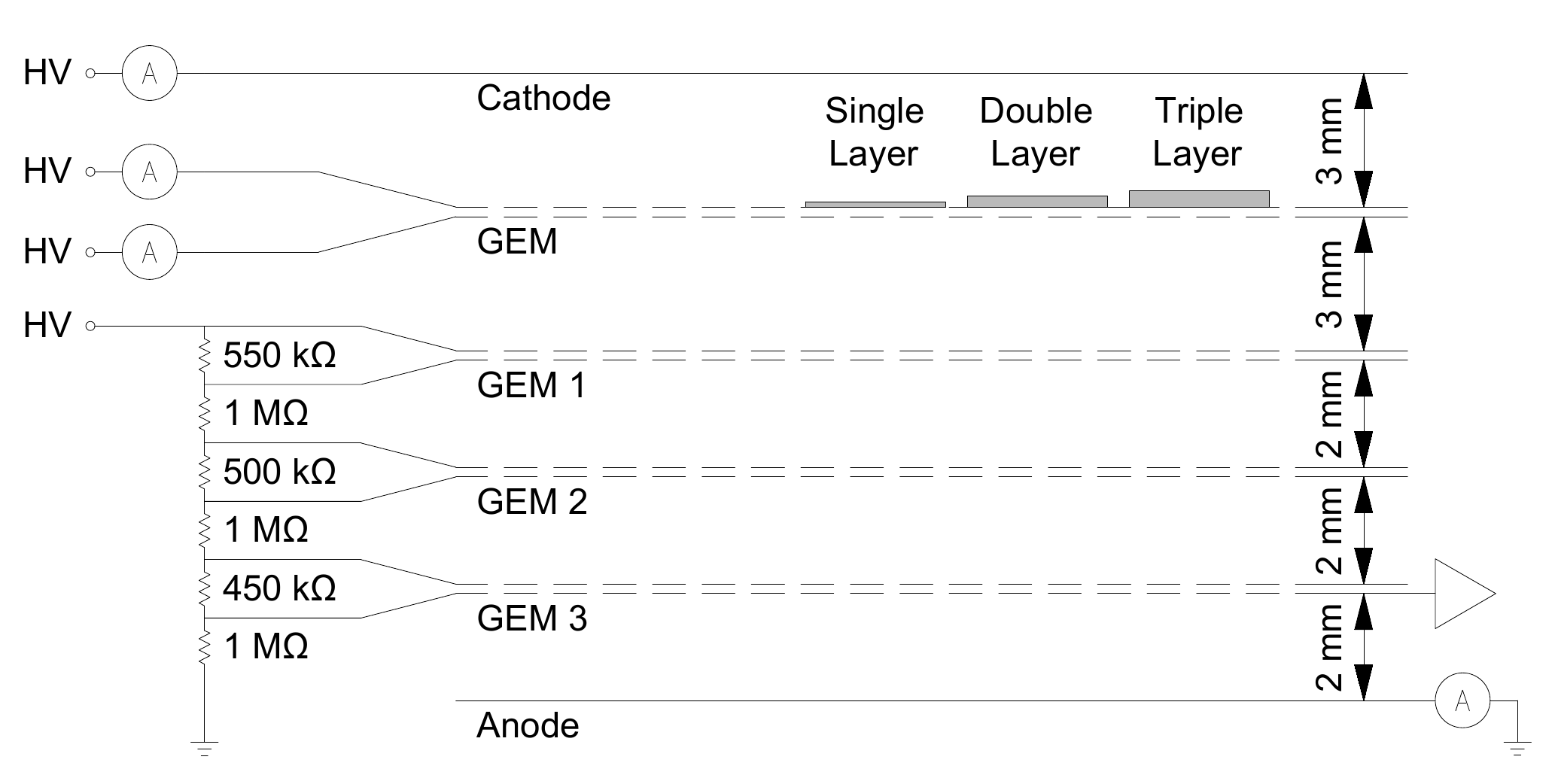}
\caption{Schematic representation of the setup. The triple GEM is used for the charge amplification, and the topmost GEM is used as support for the graphene.}
\label{fig:scheme}
\end{figure}
A schematic representation of the detector is shown in Fig.~\ref{fig:scheme}.
It consists of two 3~mm-long conversion volumes separated by a structure holding the suspended graphene.
The structure can be either a 5~$\mu$m-thick copper mesh with holes, or a GEM.
Tests of the former structures are reported in~\cite{Thuiner:2015}.
Here we report on preliminary tests of a graphene-coated GEM.
Single, double and triple graphene layers are deposited onto the GEM.
A portion of the GEM is left without graphene so that comparative measurement can be done without modifying the setup.
Despite the fact that the bulk resistivity of graphene increases significantly with the number of layers~\cite{Kuroda:2011}, the small particle flux involved in the measurement results in such small currents that the voltage changes can be neglected and any graphene layer can be considered as perfect conductor for practical purposes.
The conversion volumes are defined on top by the cathode and on the bottom by a \emph{standard} $10\times10$~cm$^2$ triple GEM stack~\cite{Altunbas:2002} powered via a resistor divider.
This configuration allows the simultaneous measurement of the currents at anode, cathode and the two electrodes of the graphene-coated GEM.
The lowest electrode of GEM~3 is connected to a preamplifier/amplifier/MCA chain for the event by event measurement of the signals.
The detector is continuously flushed with an Ar/CO$_2$~70/30 gas mixture at about 5~L/h.
Collimated X-rays of 8~keV (approximately 1~mm$^2$ beam size) generated by a copper X-ray gun are used to produce the primary ionisation charges in the conversion volumes.

Ionisation electrons from the topmost region must cross the graphene-coated GEM before undergoing multiplication in the GEMs.
Similarly, ions produced during the avalanche process in the GEMs must pass the graphene to reach the cathode.
The electron (ion) transparency $T$ is defined as the fraction of the charge from the top (bottom) volume reaching the bottom (top).
Always referring to Fig.~\ref{fig:scheme}, the primary electrons produced in the uppermost conversion volume by the X-ray conversion are suppressed by a factor $T$ when crossing the graphene, yielding a lower pulse height spectrum.
The electron transparency is defined as the ratio of the peak positions between the spectra from the events in the bottom and top conversion volumes.
For the ion transparency measurement the GEM is used as ion generator.
The ion transparency is defined as the ratio between the current at the cathode and the sum of the currents at the cathode and at the electrodes of the graphene-coated GEM.
A detailed description of the methods to extract the electron and ion transmissions can be found in~\cite{Thuiner:2015}.

\section{Discussion}
In the previous tests~\cite{Thuiner:2015} we reported a non-trivial issue regarding the micro-defects in single layer graphene, which dominate the electron and ion transparency measurements.
Although the graphene layer is almost perfect, the electrons and ions are focussed into small and sparse defects (opening in the graphene) in the same way they are focussed into the holes of a GEM.
This contribution overwhelms the one of electrons which effectively tunnel through the graphene.
To reduce the amount of defects, we prepared double and triple graphene layers transferred one on top of the other.
This resulted in a decrease of the ion transparency down to the sensitivity of the measurement, while keeping some measurable transparency to the electrons.
Without being able to exclude completely the effect of defects, we interpreted this asymmetry, though weak compared to the expected one, as a hint to the intrinsic behaviour of the graphene.

Graphene is mostly transparent to electrons with energies of the order of~keV~\cite{Meyer:2007}, but there is no common understanding of what happens to lower energies.
In addition, at lower energies, the incident angle may play a role in the destiny of the impinging electron.
The energy distribution of drifting electrons depends on the gas mixture and on the external electric field applied.
The average electron energy can be increased reducing the amount of the quencher admixture and increasing the electric field.
In Ar/CO$_2$ mixtures and at moderate electric fields this average is few~eV, but during amplification in the holes of a GEM, it exceeds 10~eV.
%For this reason the sample under study is transferred onto a GEM.

\begin{figure}
\centering
\includegraphics[width=1.0\linewidth]{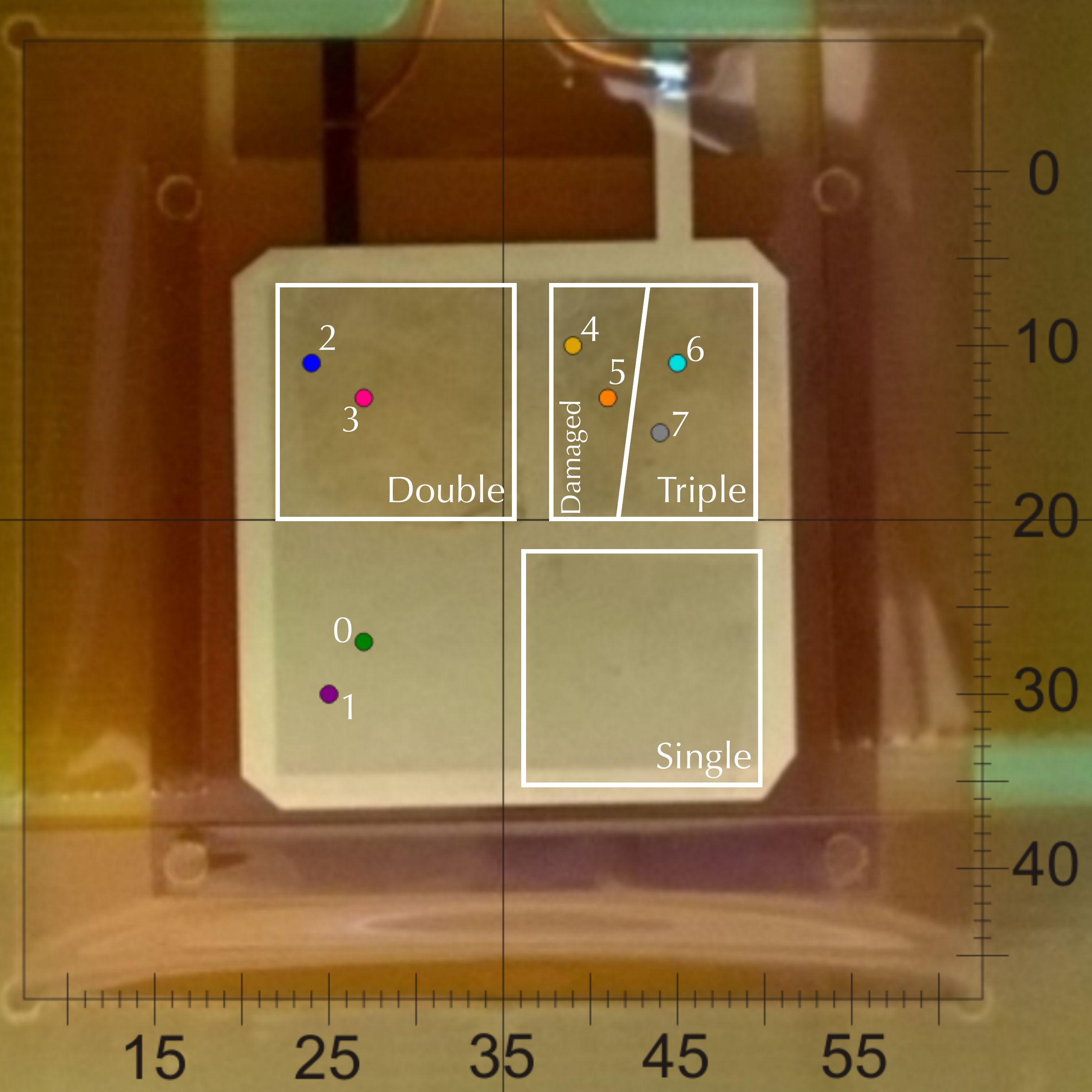}
\caption{Image of the graphene-coated GEM with dimensions in millimetre. The circles represent the position and the size of the X-ray beam. The colours and the numbering are referred to the plot in Fig~\ref{fig:trans2}.}
\label{fig:trans1}
\end{figure}

\begin{figure}
\centering
\includegraphics[width=1.0\linewidth]{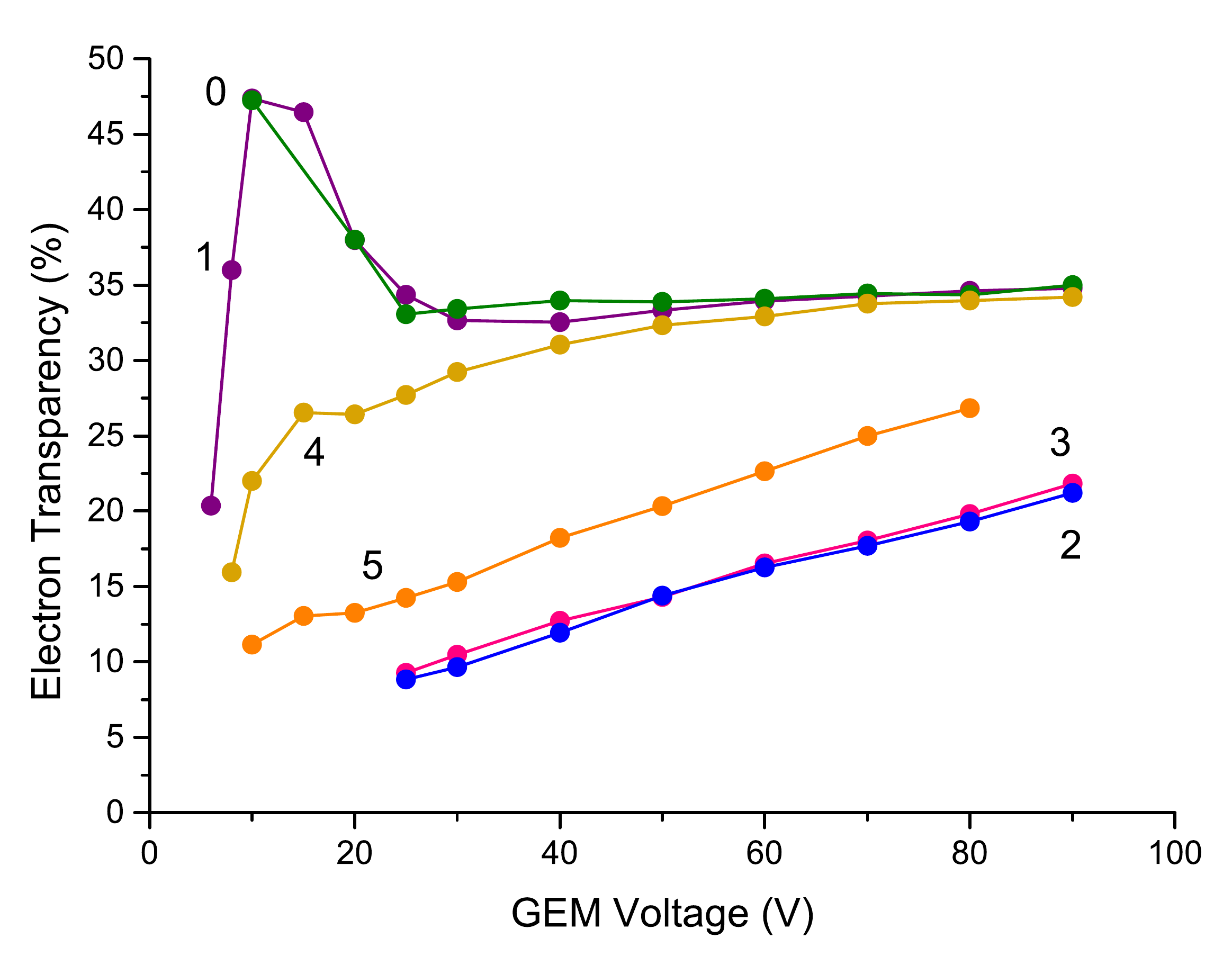}
\caption{Electron transmission through the graphene-coated GEM as a function of the voltage applied across the GEM for different X-ray beam positions. Curves~6 and~7 are omitted in this plot because compatible to zero. See text for details.}
\label{fig:trans2}
\end{figure}

The image in Fig.~\ref{fig:trans1} is a picture of the graphene-coated GEM that illustrates the size of the X-ray beam and the positions where we measured the electron transparency.
Three rectangles highlight the position of the single, double and triple layer of graphene.
The colour mapping and the numbering refer to the curves of Fig.~\ref{fig:trans2} which show the measured electron transparency as a function of the voltage across the GEM for different irradiated spots.
The nominal electric field in the top conversion volume is set to 50~V/cm, and in the bottom is 1~kV/cm.
Curves~0 and~1 are the reference ones, and they are obtained for the non-coated portion.
This transparency behaviour is very typical for GEMs~\cite{Sauli:2006}.
Curves~2 and~3 correspond to the double layer graphene.
The presence of the graphene reduces the transparency with respect to the reference curves.
The similarity of these two curves suggests that the layer is uniform.
From the SEM image, it was found that the graphene on the left side of the triple layer portion is damaged.
In fact, the curves~4 and~5 shows more transparency than the previous two.
The same SEM image showed instead no defects in the triple layer below the points~6 and~7.
The transparency in these two points is lower than the sensitivity of the measurement (few percent), and the corresponding curves in Fig.~\ref{fig:trans2} are not shown.

During the measurement of the ion transmission, a contact of few~k$\Omega$ developed between the two electrodes of the GEM.
Presumably, the short is due to a sliver of graphene from the damaged layer entering into a GEM hole.
This circumstance prevented us from applying a stable voltage across the GEM and, therefore, to accomplish the ion transparency measurement.
Despite our efforts, the sample could not be recovered.

\section{Conclusion and outlook}
In this paper we presented the latest progress regarding the application of graphene in gaseous detectors in order to eliminate the ion back-flow problem.

The improvements on the techniques for transferring the graphene from the production substrate to the final substrate allow to have an almost defect-free sample of several centimetre square.
We are exploring other methods to obtain defect-free suspended graphene layers.
In particular, we are interested in methods that do not require the transport of the graphene.
Preliminary but promising results were achieved in chemically etching holes in the copper substrate where the graphene is grown onto.

We operated a quadruple GEM detector, where the first GEM was coated with square patches of single, double and triple graphene layers.
In these conditions, we measured the electron transparency in several positions and as a function of the voltage across the GEM.
The full characterisation of ion transmission through the graphene-coated GEM will be done with new samples.

The graphene impermeability to electrons, measured also in the previous tests, may be due to intrinsic properties of graphene, or may be related to extrinsic issues.
For instance, small amount of water in the gas mixture can condense on the graphene changing its properties.
On the other hand, the production of multi-layer graphene does not guarantee that the distance between the layers is small enough so that the multi-layer graphene can be considered as a quantum-mechanic object.

Future tests will certainly involve the production of double and triple CVD graphene layers grown in a single process.
In order to disentangle the characteristics attributable to the presence of gas, we plan a measurement of the electron transmission as a function of the electron energy in vacuum.
The setup will allow to probe the graphene transparency to electrons of few~eV, and possibly as a function of their impinging angle.

%The setup consists of a vacuum chamber with a transparent window.
%Inside, a photocathode is the electron source.
%The graphene sample is positioned in front of the photocathode, and the potential difference between the photocathode and the graphene defines the electron energy.
%This setup allows to probe the graphene transparency to electrons of few~eV.
%An anode electrode is positioned behind the graphene.
%The transparency will be extracted from the measured currents at the photocathode, at the graphene, and at the anode.
%Possibly the transparency will be studied as a function of the impinging electron angle.
%Depending on the amount of current that the photocathode can deliver, the same setup may be used as well to evaluate the graphene transparency in gas.
%Finally, the full characterisation of the graphene-coated GEM will be done with a new sample.

\end{document}